\documentclass[12pt]{article}
\usepackage{epsfig,amssymb,amsmath,psfrag}

\catcode`\@=11

\newcommand{\bit}[1]{\mbox{\boldmath$#1$}}

\def \BES {} 

\def \be  {\begin{equation}}
\def \ee  {\end{equation}}
\def \ba  {\begin{eqnarray}}
\def \ea  {\end{eqnarray}}
\def \baa {\begin{eqnarray*}}
\def \eaa {\end{eqnarray*}}
\def \bb  {\begin {thebibliography} }
\def \eb  {\end{thebibliography}}
\def \lab #1 {\label{#1}}
\newcommand \ci [1] {\cite{#1}}

\newcommand\re[1]{(\ref{#1})}

\def \qqqquad {\qquad\qquad}
\def \matrix #1 {\left(\begin{array}{cc} #1 \end{array}\right)}

\def \Re {\mathop{\rm Re}\nolimits}

\def \e  {\mathop{\rm e}\nolimits}
\newcommand\lr[1]{{\left({#1}\right)}}
\newcommand \widebar [1] {\overline{#1}}

\newcommand \vev [1] {\langle{#1}\rangle}

\newcommand{\ft}[2]{{\textstyle\frac{#1}{#2}}}

\def\XXint#1#2#3{{\setbox0=\hbox{$#1{#2#3}{\int}$}
     \vcenter{\hbox{$#2#3$}}\kern-.5\wd0}}

\def \tg{\tilde{g}}

\def\numberbysection{\@addtoreset{equation}{section}
                     \def\theequation{\thesection.\arabic{equation}}}
\numberbysection

\begin{document}

\renewcommand{\thefootnote}{\fnsymbol{footnote}}

\begin{titlepage}
\begin{flushright}
\begin{tabular}{l}
LPT--Orsay--08--76 \\
ITP-Budapest Report No. 642
\end{tabular}
\end{flushright}

\vskip3cm

\begin{center}
 {\large \bf Scaling function in AdS/CFT from the O(6) sigma model }
\end{center}

\vspace{1cm}

\centerline{ \large Z. Bajnok$^{a}$, J. Balog$^{b}$, B. Basso$^{c}$, G.P. Korchemsky$^{c}$ and L.
Palla$^{d}$   } \vspace{1cm}

\begin{center}
{\sl
${}^a$ Theoretical Physics Research Group,  Hungarian Academy of Sciences

       1117 Budapest, P\'azm\'any s. 1/A, Hungary
\vspace{0.5cm}

${}^b$ Research Institute for Particle and Nuclear Physics

1525 Budapest 114, Pf. 49, Hungary

\vspace{0.5cm}

${}^c$ Laboratoire de Physique Th\'eorique, Unit\'e Mixte de Recherche du CNRS (UMR 8627)

Universit\'e de Paris XI, 91405 Orsay C\'edex, France

\vspace{0.5cm}

${}^d$    Institute for Theoretical Physics, Roland E\"otv\"os University

          1117 Budapest  P\'azm\'any s. 1/A, Hungary
}
\end{center}

\vspace{1cm}

\centerline{\bf Abstract}

\vspace{5mm}

Asymptotic behavior of the anomalous dimensions of Wilson operators with high spin and twist is
governed in planar ${\cal N}=4$ SYM theory by the scaling function which coincides at strong
coupling with the energy density of a two-dimensional bosonic O(6) sigma model. We calculate this
function by combining the two-loop correction to the energy density for the O($n$) model with
two-loop correction to the mass gap determined by the all-loop Bethe ansatz in ${\cal N}=4$ SYM
theory. The result is in agreement with the prediction coming from the thermodynamical limit of the
quantum string Bethe ansatz equations, but disagrees with the two-loop stringy corrections to the
folded spinning string solution.

\end{titlepage}

\setcounter{footnote} 0


\newpage

\renewcommand{\thefootnote}{\arabic{footnote}}

\section{Introduction}

The AdS/CFT \cite{Mal97} establishes the correspondence between Wilson operators in $\mathcal{N}=4$
super Yang-Mills theory (SYM) and states of strings spinning on $\rm AdS_5\times S^5$
background~\cite{GKP,FT}. It has been recently recognized that there exists a remarkable relation
between both theories and the two-dimensional bosonic O(6) sigma model. This relation emerges when one
studies anomalous dimensions of Wilson operators in planar $\mathcal{N}=4$ SYM theory at strong
coupling in the limit \cite{BGK06} when the Lorentz spin $N$ of the operators grows exponentially with
their twist $L$
\be\label{j}
 j=\frac{L}{\ln N} = \text{fixed}\,, \qqqquad  N,\  L\to\infty\,.
\ee
In this limit, the anomalous dimensions grow logarithmically with $N$ and their leading asymptotic
behavior is governed by the scaling function%
\footnote{More precisely, for given spin $N$ and twist $L$, anomalous dimensions of Wilson operators
occupy a band. The scaling function describes the \textit{minimal} anomalous dimension in this
band.} depending on $j$ and the t' Hooft coupling $g^2 = g_{\rm YM}^2 N_c/(4\pi)^2$. Using the dual
description of Wilson operators as folded strings spinning on ${\rm AdS_5\times S^5}$ and taking
into account the one-loop stringy corrections to these states~\ci{FTT06}, Alday and
Maldacena~\cite{AM07} conjectured that the scaling function should coincide at strong coupling with
the energy density of a two-dimensional bosonic O(6) sigma model. Recently, this relation was
established in planar $\mathcal{N}=4$ SYM theory at strong coupling \cite{BK08} using the
conjectured integrability of the dilatation operator \cite{FRS}.

The O(6) sigma-model and, in general, two-dimensional bosonic O$(n)$ sigma-models are among the best
studied field theory models. Their popularity is explained by the fact that they can be used as low
dimensional toy models of QCD: they are asymptotically free in perturbation theory, their classical
conformal invariance is broken and the mass of the physical excitations is dynamically generated by
the dimensional transmutation mechanism~\cite{PW83,FR85}. In addition they are integrable and many
physical quantities are exactly calculable~\cite{ZZ78}. These models can also be studied non-perturbatively
with a help of lattice Monte Carlo simulations making use of a very efficient simulation algorithm
(the cluster algorithm).

Using the known exact S-matrix of the O$(n)$ sigma-model, a linear thermodynamical Bethe ansatz
(TBA) integral equation can be derived describing the free energy in the presence of an external
field coupled to one of the Noether currents of the model.%
\footnote{The energy density is related to the free energy through Legendre transformation.} The
free energy can also be calculated in perturbation theory for large values of the external field due
to asymptotic freedom. The original motivation of this calculation \cite{HMN90} was that by
comparing the two results for the free energy one can calculate the physical mass $m$ of the O$(n)$
particles in terms of the perturbative $\Lambda$-parameter (dynamical scale defining solutions to
Gell-Mann-Low equation). The exact mass gap $m/\Lambda$ determined this way has been checked by
using Monte Carlo results and also by comparing finite volume mass gap values computed in
perturbation theory \cite{Shin} to those obtained from a (nonlinear) TBA integral equation
\cite{BH04}.

In this paper, we employ tools developed for the O($n$) sigma-model to compute the scaling
function in the AdS/CFT.

\subsection{Scaling function in $\mathcal{N}=4$ SYM}

The Wilson operators under consideration are built from $L$ complex scalar fields and $N$ light-cone
components of the covariant derivatives. Their minimal anomalous dimension has the following behavior
both at weak and at strong coupling~\cite{BGK06,AM07,FRS,FTT06,CK07}
\be\label{sum_up}
\gamma_{N,L}(g) = \left[ 2\Gamma_{\rm cusp}(g)+ \epsilon(g,j)\right] \ln N + \ldots\,,
\ee
where ellipses denote terms suppressed by powers of $1/L$. Here the first term inside the square
brackets has a universal, $j-$independent form and it involves the cusp anomalous
dimension~\cite{KR87,BGK03}. The dependence on the twist resides in $\epsilon(g,j)$ which is a
nontrivial function of the 't Hooft coupling and the scaling variable $j$ normalized as
$\epsilon(g,0)=0$.

At weak coupling, the scaling functions $\Gamma_{\rm cusp}(g)$ and $\epsilon(g,j)$ can be found in a
generic (supersymmetric) Yang-Mills theory in the planar limit by making use of the remarkable
property of integrability \cite{BGK06,BGK03}. In maximally supersymmetric $\mathcal{N}=4$ theory,
these functions can be determined in the planar limit for arbitrary values of the scaling parameter
$j$ and the coupling $g$ as solutions to BES/FRS equations proposed in \cite{FRS,BES}. These
equations predict the scaling function to be a bi-analytical function of $g^2$ and
$j$.~\footnote{The properties of this function at weak coupling and large $j$ were studied in
\cite{B08n}.} At strong coupling, the asymptotic expansion of the cusp anomalous dimension in powers
of $1/g$ was derived in \cite{Benna06,Alday07,Kostov07,BAF,BKK07}. It turned out that this expansion
suffers from Borel singularities \cite{BKK07} indicating that $\Gamma_{\rm cusp}(g)$ receives
nonperturbative corrections at strong coupling defined by the scale $m\sim g^{1/4} \e^{-\pi g}$
\cite{AM07,BKK07}.

At strong coupling, the scaling function $\epsilon(g,j)$ has a more complicated form and its
properties depend on a hierarchy between $g$ and $j$. As was shown in \cite{BK08}, for $g\to\infty$
and $j/m= {\rm fixed}$ (with $m$ given by \re{m_AM} below), the FRS equation for the scaling
function $\epsilon(g,j)$ can be casted into a form identical to the TBA equations for the nonlinear
$\rm O(6)$ sigma model~\cite{HMN90}. This leads to the identification of the scaling function as the
energy density $\varepsilon_{\rm O(6)}$ in the ground state of the $\rm O(6)$ model corresponding to
the particle density $\rho$
\be\label{O6-epsilon}
\varepsilon_{\rm O(6)}  = \frac{\epsilon(g,j) + j}2\,,\qqqquad \rho=\frac{j}2\,.
\ee
As was already mentioned, the $\rm O(6)$ sigma model has a nontrivial dynamics in the infrared and
it develops a mass gap. Remarkably enough, similar phenomenon also occurs for the scaling function
in $\mathcal{N}=4$ theory at strong coupling. Namely, the scaling function $\epsilon(g,j)$ depends
on a new dynamical scale~\cite{BK08,FGR08a}
\be\label{m_AM}
m = k g^{1/4} \e^{-\pi g} \left[ 1 + O(1/g)\right],\qqqquad k=\frac{2^{3/4}
\pi^{1/4}}{\Gamma(5/4)}\,,
\ee
and the same scale \re{m_AM} defines nonperturbative corrections to the cusp anomalous dimension at
strong coupling~\cite{BKK07,BK}. Later in the paper, we will compute subleading corrections to the
mass scale \re{m_AM}. The relation \re{O6-epsilon} holds for $g\to\infty$ with $j/m={\rm fixed}$ and
the scale \re{m_AM} is identified as the mass gap of the O(6) model.

\subsection{Scaling function in AdS/CFT}

The AdS/CFT correspondence relates the anomalous dimension \re{sum_up} at strong coupling to the
energy of a folded string spinning on the $\rm AdS_5\times S^5$ background~\cite{GKP,FT}
\be
\Delta=N+L+\gamma_{N,L}(g)\,,
\ee
with $N$ and $L$ being angular momenta on $\rm AdS_3$ and $\rm S^1$, respectively. Semiclassical
quantization of this stringy state yields the expansion of the anomalous dimension $\gamma_{N,L}(g)$
in powers of $1/g$. In agreement with \re{sum_up}, the coefficients of the expansion scale
logarithmically with $N$ in the limit \re{j} and determine stringy corrections to the scaling
function.

For the cusp anomalous dimension $\Gamma_{\rm cusp}(g)$, the first three coefficients of $1/g$
expansion were computed in Refs.~\cite{GKP,FTT06,Roiban:2007dq}. As a nontrivial test of the AdS/CFT
correspondence, they were found to be the same as in the $\mathcal{N}=4$ SYM theory~\cite{BKK07}.%
\footnote{Note that the semiclassical approach does not take into account nonperturbative
corrections to $\Gamma_{\rm cusp}(g)$.} For the scaling function similar calculation was performed
by two different approaches -- from two-loop stringy corrections to the folded spinning string
solution~\cite{RT07} and from thermodynamical limit of quantum string Bethe ansatz
equations~\cite{CK07,B08,G08}, leading to (in notations of \cite{RT07})
\be\label{e-2-loop}
\epsilon(g,j)+j = 2\ell^2\left[ g +\frac{1}{\pi}\left(\frac{3}{4}-\ln{\ell}\right) + \frac{1}{4\pi^2
g}\left(\frac{q_{02}}2-3\ln{\ell}+4(\ln{\ell})^2\right) +
\mathcal{O}\left({1}/{g^2}\right)\right]+O(\ell^4)\,,
\ee
with $\ell = j/(4g)$. This relation is valid at strong coupling for $j\ll g$, or equivalently
$\ell\ll 1$. Here the first two terms inside the square brackets describe, correspondingly, the
classical expression and one-loop correction to the scaling function $\epsilon(g,j)$. The last term
describes the two-loop correction and it depends on a constant $q_{02}$. The two approaches
mentioned above produce two different values of $q_{02}$
\be\label{discrepancy}
 q_{02}\bigg|_{{\rm \tiny ref\!.\cite{RT07}}} = - 2 \textrm{K}-\frac32\ln 2 +\frac 74\,,\qqqquad
 q_{02}\bigg|_{{\rm \tiny ref\!.\cite{G08}}} = -\frac32\ln 2 +\frac{11}4\,,
\ee
with $\textrm{K}$ being the Catalan constant. The two results agree with each other in term $\sim
\ln 2$ but disagree in the rest. The reason for this discrepancy remains unclear.

The semiclassical approach allows us to calculate the scaling function \re{e-2-loop} in the form of
a double series in $1/g$ and $\ell^2$. It does not take however into account nonperturbative
corrections to the scaling function  which are exponentially small as $g\to\infty$.
Alday and Maldacena~\cite{AM07} put forward the proposal that the
scaling function $\epsilon(g,j)$ can be found {\em exactly} at strong coupling in the limit $j\ll g$
and $j/m={\rm fixed}$ (with  $m$ defined in \re{m_AM}). They argued that quantum corrections in the
$\rm AdS_5\times S^5$ sigma model are dominated in this limit by the contribution of massless
excitations on $\rm S^5$ whose dynamics is described by a (noncritical) two-dimensional bosonic $\rm
O(6)$ sigma-model equipped with a UV cut-off determined by the mass of massive excitations. In terms
of parameters of the underlying $\rm AdS_5\times S^5$ sigma model, the exact value of the mass gap
is given by \re{m_AM}. The dependence of the mass scale \re{m_AM} on the coupling constant is fixed
by the two-loop beta-function of the $\rm O(6)$ model whereas the prefactor $k$ was determined in
\cite{AM07} by matching the first two terms of the semiclassical expansion \re{e-2-loop} into known
one-loop perturbative correction to the energy density of the O(6) model.

The subleading $O(1/g)$ correction to the scaling function \re{e-2-loop} involves both constant term
$q_{02}$ and logarithmically enhanced terms. As was shown in \cite{AM07}, the latter terms are
controlled by renormalization group. It is the constant $q_{02}$ that lies at crux of the relation
\re{e-2-loop} to two-loop order. The relation \re{O6-epsilon} allows us to determine this constant
by computing corrections to the energy density of the O(6) model and by matching the resulting
expression for the scaling function into \re{e-2-loop}.

\subsection{Two-dimensional O(6) sigma model}

Substitution of \re{e-2-loop} into relation \re{O6-epsilon} yields a definite prediction for the
energy density of the O(6) model in the regime of large particle density $\rho \gg m$. In this
regime the model is known to be weakly coupled and the energy density can be computed
perturbatively.

The two-dimensional O(6) sigma model is an exactly solvable quantum field theory. It is
asymptotically free at short distances, while in the infrared it develops a mass gap. The massive
excitations form the vector multiplet of the O(6) group and their S-matrix has a factorized form
\cite{ZZ78}. This makes it possible to calculate the mass gap $m$ in terms of the parameter
$\Lambda$. The idea is to consider the O(6) model in the presence of an external field $h$ coupled
to one of the conserved charges, say $Q^{12}$ (see Eq.~\re{Q} below), and calculate the change in
the free energy density in two different ways: from thermodynamical Bethe ansatz and from
perturbative expansion \cite{HMN90}.

If the external field exceeds the mass gap, $h\ge m$, a finite density of particles $\rho$ is formed
in the ground state. The corresponding ground-state energy density $\varepsilon_{\rm O(6)}(\rho)$ is
given by
\begin{equation}\label{tba2}
\varepsilon_{\rm O(6)}=m\int_{-B}^{B}\frac{d\theta}{2\pi}\,\chi(\theta)\cosh\theta\,,
\qquad\rho=\int_{-B}^{B}\frac{d\theta}{2\pi}\,\chi(\theta)\,,
\end{equation}
where the rapidity distribution satisfies an integral TBA equation
\begin{equation}\label{tba1}
\chi(\theta)=\int_{-B}^{B}d\theta^{'}K(\theta-\theta^{'}) \chi(\theta^{'})+m\cosh\theta\,.
\end{equation}
Here the kernel $K(\theta)=\lr{\log S(\theta)}'/{(2\pi i)}$ is given by the logarithmic derivative of
the scattering matrix of the particles corresponding to the largest eigenvalue of the charge
$Q^{12}$
\be
K(\theta)=\frac{1}{4\pi^{2}}\left[\psi\left(1+\frac{i\theta}{2\pi}\right)
+\psi\left(1-\frac{i\theta}{2\pi}\right)-\psi\left(\frac{1}{2}
-\frac{i\theta}{2\pi}\right)-\psi\left(\frac{1}{2}
+\frac{i\theta}{2\pi}\right)+\frac{2\pi}{\cosh\theta}\right]\,,
\ee
where $\psi(x)=\Gamma'(x)/\Gamma(x)$ denotes the logarithmic derivative of the gamma function.

The free energy density $\mathcal{F}(h)$ can be obtained from \re{tba2} through the Legendre
transformation, $\mathcal{F}(h) = \min_{\rho} \left[\varepsilon(\rho)-h\rho \right]$. Due to
asymptotic freedom, for $h\gg m$ the change in the free energy density can also be calculated
perturbatively with the result
\begin{equation}
\mathcal{F}(h)-\mathcal{F}(0)=-\frac{h^{2}}{\pi}\left\{ \frac{1}{\alpha}
-\frac{1}{2}-\frac{\alpha}{8}+O(\alpha^{2})\right\}, \label{F1}
\end{equation}
where the coupling $\alpha=\alpha(h)$ is defined in a renormalization group invariant way by
\begin{equation}\label{aldef1}
\frac{1}{\alpha}+\frac{1}{4}\ln\alpha=\ln\frac{h}{\Lambda_{\overline{\rm MS}}}\,,
\end{equation}
with $\Lambda_{\overline{\rm MS}}$ being the $\Lambda$ scale in the $\overline{\rm MS}$ scheme. The
first two terms in the right-hand side of \re{F1} were found in \cite{HMN90} while calculation of
the coefficient in front of $\alpha$ is one of the main results of this paper.

At large $h$, the solution to the TBA equations \re{tba2} and \re{tba1} can be constructed using the
generalized Wiener-Hopf technique \cite{HMN90}. The result for $\mathcal{F}(h)-\mathcal{F}(0)$ is
given by series in  $[\ln(h/m)]^{-1}$. The calculation is rather involved and only the first few
terms have been calculated so far \cite{HMN90}. Matching these terms into perturbative expansion
\re{F1}, one can establish the relation \cite{HMN90} between the mass gap $m$ and the scale
$\Lambda_{\overline{\rm MS}}$
\be\label{zeta}
\zeta=\ln\frac{m}{\Lambda_{\overline{\rm MS}}}=\frac{1}{4}(3\ln 2-1)-\ln\Gamma
\bigl(\ft{5}{4}\bigr)\,.
\ee
Taking into account this relation, we find from \re{F1} and \re{aldef1} the following expression for
the ground state energy density as a function of $\rho/m$
\be\label{eee}
\varepsilon_{\rm O(6)} =\frac{\rho^{2}\tilde{\alpha}\pi}{4}\left\{ 1
+\frac{\tilde{\alpha}}{2}+\frac{\tilde{\alpha}^{2}}{8}+O(\tilde{\alpha}^{3})\right\}\,,
\ee
where $\tilde{\alpha}=\tilde\alpha(\rho)$ denotes another useful coupling
\be
\frac{1}{\tilde{\alpha}} -\frac{3}{4}\ln\tilde{\alpha}=\ln\frac{\rho}{m}+\ln\frac{\pi}{2}+\zeta\,.
\ee
The relation \re{eee} defines the energy density $\varepsilon_{\rm O(6)}$ as a function of the
particle density $\rho$ and the mass gap $m$. We can now apply \re{O6-epsilon} to translate it into
the dependence of the scaling function $\epsilon(g,j)$ on $j$ and $m$ in the limit $m\ll j\ll g$.

To obtain the strong coupling expansion of $\epsilon(g,j)$ from \re{O6-epsilon} we also need the
explicit form of the function $m=m(g)$. To leading order it is given by \re{m_AM} while the
subleading correction to $m$ will be computed below (see  Eq.~\re{q02}). Replacing the mass scale
$m$ in \re{eee} by its explicit expression \re{m_AM}, we find that the scaling function admits the
same perturbative expansion as \re{e-2-loop} and provides a definite prediction for the constant
$q_{02}$. This constant depends on subleading corrections both to the energy density
$\varepsilon_{\rm O(6)}$ and to the mass scale $m$. We find that with these corrections taken into
account the relation \re{O6-epsilon} leads to $q_{02}=-3\ln 2/2+{11}/4$ (see Eq.~\re{discrepancy}),
in agreement with the quantum string Bethe ansatz result of \cite{G08}.

The paper is organized as follows. In Sect.~2 we summarize the properties of the scaling function in
planar $\mathcal{N}=4$ SYM at strong coupling. We show that for $m \ll j \ll g$ the scaling function
has the same form as \re{e-2-loop} and obtain the expression for the constant $q_{02}$ which
involves corrections both to the mass gap $m$ and to the energy density $\varepsilon_{\rm O(6)}$.
Then, we compute subleading $O(1/g)$ correction to the mass scale \re{m_AM}. In Sect.~3, we
calculate the ground-state energy density of the O($n$) sigma model for large particle density
$\rho$ using perturbation theory technique. We obtain an expression for the energy density
$\varepsilon_{{\rm O}(n)}$ to second order in the effective coupling constant expansion and
demonstrate that it is in agreement with numerical solution of the TBA equations. Finally, we use
the obtained expressions for $m$ and $\varepsilon_{{\rm O}(6)}$ to compute the constant $q_{02}$.
Section 4 contains our concluding remarks. Some technical details are presented in two Appendices.

\section{Scaling function in $\mathcal{N}=4$ SYM at strong coupling}

For $g\to\infty$ and $j/m={\rm fixed}$, the scaling function $\epsilon(g,j)$ is related to the
energy density of the O(6) model through relation \re{O6-epsilon}. To compute $\epsilon(g,j)$ from
\re{O6-epsilon} we have to accomplish two tasks:  (i) to determine the energy density
$\varepsilon_{\rm O(6)}$ as a function of particle density $\rho=j/2$ and mass gap $m$, and (ii)
calculate the mass scale $m$ from the FRS equation as a function of the coupling constant $g$.

The leading order expression for the mass scale is given by \re{m_AM}.  The subleading correction
to $m$ can be parameterized as follows
\be\label{m1}
m = k g^{1/4} \e^{-\pi g} \left[ 1 +\frac{m_1}{\pi g}+ O(1/g^2)\right],
\ee
with $k$ defined in \re{m_AM} and $g-$independent parameter $m_1$ to be determined. For the energy
density  $\varepsilon_{\rm O(6)}$, we make use of the results of previous studies of the TBA
equations for the two-dimensional O($n$) model~\cite{HMN90}. For arbitrary $j/m$, solutions to the
TBA equations \re{tba2} and \re{tba1} do not admit a simple analytical representation. However, they
can be constructed in the limits $j/m \ll 1$ and $j/m\gg 1$, which correspond, respectively, to
(nonperturbative) small and (perturbative) large particle density regimes:
\begin{itemize}
\item
For $j\ll m$, the integral equations \re{tba2} and \re{tba1} can be solved by iterations leading
to~\cite{HMN90,AM07,BK08}
\be\label{11}
\epsilon(j,g) +j  =  m^2\left[\frac{j}{m} + \frac{\pi^2}{24} \lr{\frac{j}{m}}^3+
\ldots\right],
\ee
where expansion runs in powers of $j/m$. This regime was extensively studied both numerically and
analytically and subleading corrections to \re{11} were recently computed in \cite{FGR08a}.

\item For $j\gg m$, the expression for the scaling function follows from the known perturbative
result for the energy density of ${\rm O}(6)$ model~\cite{HMN90,AM07,BK08}
\begin{align}\label{e-pt}
\epsilon(j,g) +j  =  \frac{\pi j^2}{8\ln(j/m)} \left[1+ \frac34{\frac{\ln\lr{\kappa
\ln(j/m)}+\ft12}{\ln(j/m)}}+\frac{9}{16} {\frac{\ln^2(\kappa\ln(j/m))+\epsilon_1}{\ln^2(j/m)}}+\ldots\right],
\end{align}
where $\ln \kappa = \frac12 -\frac13\ln2-\frac43\ln\Gamma(\ft34)$ and the constant
$\epsilon_1$ remains unknown.
\end{itemize}
The relation \re{e-pt} resums through renormalization group (an infinite number of) perturbative
corrections in $1/g$ which are proportional to $j^2$ and are enhanced by powers of
$\ln(j/g)$~\cite{AM07,RT07}. Replacing the mass scale in \re{e-pt} by its expression \re{m1} and
re-expanding the right-hand side of \re{e-pt} in powers of $1/g$, we arrive at the relation
\re{e-2-loop} and obtain the constant $q_{02}$ as
\be\label{q02}
q_{02} = \frac98 + 8 m_1 + \frac92 \epsilon_1\,.
\ee
It depends on the parameters $m_1$ and $\epsilon_1$ entering \re{m1} and \re{e-pt}, respectively.
In this section, we compute $m_1$ and return to  $\epsilon_1$ in section~3.

\subsection{Mass scale}

The dependence of the mass scale $m$ on the coupling follows univocally from the FRS equation. As
was shown in \cite{BK08}, it has the following form
\be\label{m=int}
m =  \frac{8\sqrt{2}}{\pi^2}\e^{-\pi g}-\frac{8g}{\pi}\e^{-\pi g}\Re\left[ \int_0^\infty \frac{dt\,
\e^{i(t-\pi/4)}}{t+i\pi g}\left(\Gamma^{\BES}_{+}(t)+ i \Gamma^{\BES}_{-}(t)\right)\right],
\ee
where $\Gamma^{\BES}_{\pm}(t)$ are real functions of $t$ also depending on the coupling constant but
independent on the scaling variable $j$. In addition, these functions have a definite parity,
$\Gamma^{\BES}_{\pm}(-t) = \pm \Gamma^{\BES}_{\pm}(t)$, and satisfy an integral equation
\cite{BKK07} which follows from the BES equation \cite{BES}. To save space, we do not present it
here and refer an interested reader to \cite{BK08,BKK07} for details.

It is interesting to note that the  functions  $\Gamma_\pm^{\BES}(t)$ also play a distinguished role
in determination of the cusp anomalous dimension in planar $\mathcal{N}=4$ SYM theory.  Namely,  for
arbitrary coupling $\Gamma_{\rm cusp}(g)$ can be derived from the asymptotic behavior of
$\Gamma_\pm^{\BES}(t)$  at small~$t$
\be\label{cusp}
\Gamma_{\rm cusp}(g) = - 2g \left(\Gamma^{\BES}_{+}(0)+ i \Gamma^{\BES}_{-}(0)\right).
\ee
This relation was used in \cite{BKK07} to obtain the asymptotic expansion of $\Gamma_{\rm cusp}(g)$
in powers of $1/g$.

The functions $\Gamma_\pm(t)$ were constructed in \cite{BKK07} in the form of Neumann series over
Bessel functions 
\ba\label{G-sol}
\Gamma_+(t) &=& \sum_{k=0}^\infty (-1)^{k+1}   J_{2k}(t)\Gamma_{2k}(g)\,,
\\\nonumber
\Gamma_-(t) &=& \sum_{k=0}^\infty (-1)^{k+1}   J_{2k-1}(t)\Gamma_{2k-1}(g)\,.
\ea
Here the $g-$dependent expansion coefficients are given by
\be\label{Gam}
 \Gamma_{-1}=1\,,\qquad
\Gamma_{k} (g)=  -\frac12\Gamma_{k}^{(0)}+\sum_{p=1}^\infty \frac1{g^p} \left[c_p^-(g)
\Gamma_{k}^{(2p-1)} + c_p^+ (g) \Gamma_{k}^{(2p)} \right]\,, \quad (k\ge 0)\,,
\ee
where the dependence on $k$ is carried by the coefficients $\Gamma_{k}^{(p)} $ defined as
\be\label{Gam-k}
\Gamma_{2k}^{(p)} = \frac{\Gamma(k+p-\ft12)}{\Gamma(k+1)\Gamma(\ft12)}, \qquad
\Gamma_{2k-1}^{(p)} = (-1)^{p}\frac{\Gamma(k-\ft12)}{\Gamma(k+1-p)\Gamma(\ft12)}\,,
\ee
while the dependence on the coupling resides in $c_p^\pm(g)$.

The functions $\Gamma_\pm(t)$ defined in \re{G-sol} are uniquely specified by the expansion
coefficients $c_p^\pm(g)$. The latter satisfy the quantization conditions~\cite{BKK07}
\begin{align}\label{QC}
& \sum_{p\ge 0} (2\pi s)^{p}\,c_p^+ (g) \frac{\Gamma(p-\ft14)}{2\Gamma(\ft34)} =
\frac{\Gamma(\ft34)\Gamma \left( 1- s\right)}{\Gamma  \left( \ft34- s \right) } + O(1/g)\,,
\\ \notag
& \sum_{p\ge 0} (2\pi s)^{p}\, \left[c_p^-(g) +
c_p^+(g)\lr{2p-\ft32}\lr{p-\ft14}\right]\frac{4\Gamma(p-\ft34)}{\Gamma(\ft14)} =
\frac{\Gamma(\ft14)\Gamma \left( 1-s \right) }{\Gamma \left( \ft14-s \right)}+ O(1/g)\,,
\end{align}
with $c_0^+ = -\frac12$, $c_0^- = 0 $ and $s$ being arbitrary. Comparing
coefficients in front of powers of $s$ in both sides of these relations, we can determine
$c_p^\pm(g)$ in  the form of asymptotic series in $1/g$. In this manner, we get
\begin{align} \notag
c_1^+(g) & = -3\,{\frac {\ln 2 }{\pi }}+\frac12+ O(1/g)\,,\qquad  \\ c_1^-(g) &= \frac34\,{\frac
{\ln 2 }{\pi }}-\frac14+ O(1/g)\,,\quad \ldots
\end{align}
Substituting these relations into \re{G-sol} and \re{Gam} and performing
summation over $k$
in the right-hand side of \re{Gam} we obtain after some algebra
\begin{align}\label{sce}
 \Gamma^{\BES}_{+}(it)+i \Gamma^{\BES}_{-}(it)  =-V_0(t)
  -(4\pi g)^{-1}\left[ \left(\frac{\pi}{2}-3\ln{2}\right)t V_0(t)
- \frac{3\ln{2}}{2} V_1(t)\right] +\mathcal{O}\left(1/g^2\right),
\end{align}
where the notation was introduced for the functions
\begin{align} \label{II}
V_0(t)&=\frac{\sqrt{2}}{\pi}\int_{-1}^1 du\e^{t u} \lr{\frac{1+u}{1-u}}^{1/4}\,,
\\ \notag
V_1(t)&=\frac{\sqrt{2}}{\pi}\int_{-1}^1 du\e^{t u} \lr{\frac{1+u}{1-u}}^{1/4}\frac1{1+u}\,.
\end{align}
The first term in the right-hand side of \re{sce} coincides with the leading-order solution found in
\cite{Alday07}. Moreover, taking $t=0$ in \re{sce} we verify using \re{cusp} that
\be
\Gamma_{\rm cusp}(g)/(2g)=- {\Gamma_+^{\BES}(0) - i\Gamma_-^{\BES}(0)}=  1  - \frac{3\ln 2}{4\pi g} + \ldots\,,
\ee
in agreement with the known strong coupling expansion of the cusp anomalous dimension.

\subsection{Correction to the mass gap}

Let us now apply \re{m=int} and \re{G-sol} to compute the mass scale at strong coupling. We observe
that the $t-$integral in the right-hand side of \re{m=int} receives a dominant contribution from the
region $t\sim g$. Trying to apply \re{sce} we recognize that the expansion \re{sce} is not
well-defined in this region because, due to the presence of $t$ inside square brackets, expansion
parameter is $t/(4\pi g) = O(g^0)$. Thus, in order to compute the mass scale from \re{m=int}, we
have to resum the whole series \re{sce}  in the double scaling limit $t, g \rightarrow \infty$ with
$t/g=\rm fixed$.  Fortunately, this particular limit was already studied in \cite{BKK07}.

In the double scaling limit, it is convenient to change the integration variable in \re{m=int} as
$t\to 4\pi g i t$ and expand the function $\Gamma_\pm^{\BES}(4\pi g i t)$ into
series in $1/g$
with $t=O(g^0)$. We find from \re{sce} that the expansion has the following form
\begin{align}\label{rsce}
\Gamma^{\BES}_{+}(4\pi g it)+ i \Gamma^{\BES}_{-}(4\pi g it) &= f_{0}(t)  V_0(4\pi g t) + f_{1}(t)
V_1(4\pi g t)\,,
\end{align}
with $f_0(t)  = - 1- \left(\frac{\pi}{2}-3\ln{2}\right)t + O(t^2,1/g)$ and $ f_1(t) = O(1/g)$. To
determine the functions $f_0(t)$ and $f_1(t)$, we solve the quantization conditions \re{QC} and,
then, use  the obtained expressions for $c_p^\pm(g)$ to resum the series in \re{Gam}.  In this way,
we obtain that $f_0(t)$ and $f_1(t)$ are given by a linear combination of the ratio of Euler
gamma-functions (see Eq.~\re{ff} in Appendix~A). Their substitution into \re{rsce} yields the
expression for functions $\Gamma^{\BES}_{\pm}(4\pi g it)$ in the double scaling limit~\cite{BK}
\begin{align}\label{G-final}
 \Gamma^{\BES}_{+}(4\pi g it)+ i \Gamma^{\BES}_{-}(4\pi g it)  = & - V_0(4\pi gt)
\frac{{\Gamma}(\ft34)\Gamma(1-t)}{\Gamma(\ft34-t)}
\\[2mm] \notag
& + \frac{V_1(4\pi gt)}{4\pi
g}\left[\frac{\Gamma(\ft14)\Gamma(1+t)}{4t\,\Gamma(\ft14+t)}-\frac{\Gamma(\ft34)\Gamma(1-t)}{4t\,\Gamma(\ft34-t)}
\right]
\\[3mm] \notag   &   + \frac{V_0(4\pi gt)}{4\pi g} \left[\lr{\frac{3\ln 2}{4}+\frac1{8t}}
\frac{\Gamma(\ft34)\Gamma(1-t)}{\Gamma(\ft34-t)}- \frac{\Gamma(\ft14)\Gamma(1+t)}{8t\,
\Gamma(\ft14+t)} \right]\bigg\}
  +\ldots\,,
\end{align}
where ellipses denote subleading terms suppressed by powers of $1/g$  and the functions $V_0(4\pi
gt)$ and $V_1(4\pi gt)$ are defined in \re{II}. The relation \re{G-final} is consistent with the
expansion \re{sce} and it can be used to compute the $1/g$ correction to the mass gap \re{m=int}.

We are now ready to compute the mass scale \re{m=int}. To this end, we substitute \re{G-final} into
\re{m=int} and work out the asymptotic expansion of the $t-$integral at large $g$. For the first
term in the right-hand side of \re{G-final} this calculation was already performed in \cite{BK08}.
It leads to the expression for $m$ which coincides with the mass gap \re{m_AM} found from the string
theory considerations \cite{AM07}. Taking into account the remaining terms in the right-hand side of
\re{G-final} we should be able to compute the subleading correction to $m$. Calculation goes along
the same lines as in \cite{BK08} and it leads to (see Appendix~A for more details)
\begin{align} \label{m}
m & = k g^{1/4} \e^{-\pi g}\left[1 +\frac{3-6\ln 2}{32\pi g} +O(1/g^2) \right]\,,
\end{align}
with $k={2^{3/4} \pi^{1/4}}/{\Gamma(5/4)}$. Comparing this relation with \re{m1} we conclude that
\be\label{megy}
m_1 = \frac{3}{32}-\frac{3}{16}\ln 2\,.
\ee
This result is in an agreement with the numerical value found in the last reference in
\cite{FGR08a}.

The coefficient in front of $(32\pi g)^{-1}$ in \re{m} is given by the sum of two terms of different
transcendentality. We observe that $\sim\ln 2$ term can be absorbed into redefinition of the
coupling constant
\be
 g' = g - \frac{3\ln 2}{4\pi}\,,
\ee
so that the mass gap \re{m} in terms of the shifted coupling $g'$ looks as
\be
m =  \frac{(\pi g')^{1/4} \e^{-\pi g'}}{\Gamma(5/4)}\left[1 +\frac{3}{32\pi g'} + O(1/g'^2)
\right]\,.
\ee
It is interesting to notice that similar simplification also occurs for the cusp anomalous dimension
$\Gamma_{\rm cusp}(g)$ at strong coupling. As was found in \cite{BKK07}, the expansion coefficients
in the strong coupling expansion of $\Gamma_{\rm cusp}(g)$ also involve terms $\sim\ln 2$ but they
disappear after one re-expands the series in $1/g'$. This suggests that the expansion parameter at
strong coupling is $g'$ rather than $g$.

\subsection{Induced renormalization scheme}

As was already mentioned, the mass scale $m$ emerges in the AdS/CFT correspondence through
dimensional transmutation mechanism  in an effective two-dimensional theory describing dynamics of
massless excitations in the ${\rm AdS_5\times S^5}$ sigma model.  As a result, the dependence of the
mass scale of the coupling constant is dictated by the renormalization group.

The coupling constant in the effective theory depends on the scale and is related to the coupling of
the (conformal invariant) ${\rm AdS_5\times S^5}$ sigma model as $\bar g^2 (\mu)=1/(2g)$ with the
scale $\mu\sim 1$ defined by masses of massive excitations~\cite{AM07,RT07}. The coupling $\bar g
(\mu)$ satisfies the Gell-Man--Low equation
\be\label{Gell}
\mu\frac{d\bar g}{d\mu} = \beta(\bar g) = - \beta_0 \bar g^3 -\beta_1 \bar g^5 -\beta_2\bar g^7
+O(\bar g^9)\,,
\ee
and it leads to the following expression for a renormalization group invariant scale
\be\label{Lam}
\Lambda = \mu \e^{-\int^{\bar g} \frac{d\bar g}{\beta(\bar g)} } = \mu\e^{-\frac1{2\beta_0\bar
g^2}}\bar g^{-\beta_1/\beta_0^2}\left[1+
\frac1{2\beta_0}\lr{\frac{\beta_1^2}{\beta_0^2}-\frac{\beta_2}{\beta_0}}\bar g^2 + O(\bar g^4)
\right].
\ee
Then, the relation \re{O6-epsilon} between the scaling function and O(6) sigma model implies that
$\beta(\bar g)$ in \re{Gell} should coincide with the beta-function of this model. In bosonic
two-dimensional O(6) sigma model the beta-function coefficients are given by \cite{BH}
\be\label{beta}
\beta_0 = \frac{1}{\pi}\,,\qqqquad \beta_1 = \frac{1}{2\pi^2}\,.
\ee
Notice that the beta function \re{Gell} depends on the renormalization scheme starting from $O(\bar
g^7)$ term. The same is true for the scale $\Lambda$ while the mass scale $m$ should be scheme
independent. The two scales are related to each other as
\be\label{m=Lambda}
m = c\, \Lambda\,,
\ee
where the coupling independent factor $c$ is needed to restore the scheme independence of $m$. In
the special case of the $\widebar{\rm MS}$ scheme, this relation takes the form \re{zeta} with
$c_{\widebar{\rm MS}} = \e^{\zeta}$.

Replacing the beta-function coefficients in \re{Lam} by their actual values \re{beta} we obtain from
\re{m=Lambda}
\begin{align}
m =  c\, \mu \e^{-\frac{\pi}{2\bar g^2}}\bar g^{-1/2}\left[1+
 \lr{\frac{1}{8\pi}-\frac{\pi^2}2{\beta_2}}\bar g^2 + O(\bar g^4) \right].
\end{align}
Let us now compare this relation with the expression for the mass scale \re{m}
obtained from exact solution of the FRS equation. We find that the two
expressions indeed coincide upon identification
\be
\bar g^2(\mu)=\frac1{2g}\,,\qquad  \qquad  c^{\rm (FRS)}\,
  \mu =  
  \frac{2^{1/2}\pi^{1/4}}{\Gamma(5/4)}\,,\qqqquad \beta_2^{\rm (FRS)} = \frac1{8\pi^3} \left(6\ln 2 -1 \right)\,.
  \ee
Here we introduced the subscript (FRS) to indicate that these expressions are valid
in a particular renormalization scheme dictated by the FRS equation.

\section{Energy density in the two-dimensional O($n$) sigma model}

The aim of this section is to calculate the ground-state energy density of the O($n$) sigma model
for large particle density $\rho$. The ground-state energy density $\varepsilon_{{\rm O}(n)}$ can be
obtained either from the solution of the integral TBA equations, Eqs.~(\ref{tba1}) and (\ref{tba2}),
or from perturbation theory. To compute the constant \re{q02} and, then, to make a comparison with
the results of Refs.~\cite{RT07} and \cite{G08}, one needs the second subleading correction to
$\varepsilon_{\rm O(6)}$ in the large $\rho$ limit analogous to $\epsilon_1$ in  \re{e-pt}. Since
the techniques on the TBA side are not developed enough to perform the expansion at such depth, we
determine this correction using standard perturbation theory.

\subsection{Perturbative calculation of the free energy density}

The fundamental fields of the O($n$) nonlinear sigma model are $\bit{S}(x)=(S^1,\ldots,S^n)$ subject
to the constraint $\sum_1^n S^{j}S^{j}=1$. The theory has global O($n$) symmetry and the corresponding
conserved charges can be written as
\be\label{Q}
Q^{ij}=\int J_{0}^{ij}dx_1\,,\quad\qquad
J_{\mu}^{ij}=S^{i}\partial_{\mu}S^{j}-S^{j}\partial_{\mu}S^{i}\,.
\ee
We couple an external field to the conserved charge $Q^{12}$ and define the theory by its Euclidean
two-dimensional Lagrangian:
\be\label{L}
\mathcal{L}(x)=\frac{1}{2\lambda^{2}}\left[ \partial_\mu \bit{S}\,\partial_\mu \bit{S} +2ih(S^{1}
\partial_{0}S^{2}-S^{2}\partial_{0}S^{1})+h^{2}\left\{ (S^{3})^{2}
+\dots+(S^{n})^{2}-1\right\} -2\omega^{2}S^{1}\right]\,,
\ee
where $\lambda$ is the bare coupling constant and $h-$dependent terms are chosen in such a way that they
modify the Hamiltonian of the model by term $(-h Q^{12})$.

In order to avoid infrared divergences we introduced in the right-hand side of \re{L} an extra term
with regulator $\omega$ which is going to be put to zero at the end of the calculation. This extra
term fixes the classical ground-state to be
\be
S^{1}=1\,, \qquad S^{2}=S^{3}=\dots=S^{n}=0\,.
\ee
We parameterize the small fluctuations around this ground-state by exploiting the remaining
symmetries:
\be\label{S-z}
S^{1}=\sqrt{1-\lambda^{2}(y^{2}+\bit{z}^{2})}\,,\quad\qquad S^{2}=\lambda y, \quad S^{3}=\lambda
z^{1},\quad\dots\,,\quad S^{n}=\lambda z^{n-2}\,,
\ee
where the fields $\bit{z}=(z^{1},\dots,z^{n-2})$ form the vector representation of the unbroken
O$(n-2)$. We substitute \re{S-z} into \re{L} and expand the Lagrangian to second order in the
coupling $\lambda$ to get
\be\label{L-dec}
\mathcal{L}(x)=\lambda^{-2}\mathcal{L}_{-2}+\mathcal{L}_{0}+\lambda\mathcal{L}_{1}
+\lambda^{2}\mathcal{L}_{2}+O(\lambda^{3})\,,
\ee
where the various terms depend on the parameter $M^{2}=h^{2}+\omega^{2}$ and look as follows
\begin{eqnarray}\label{L's}
\mathcal{L}_{-2} & = & -\frac{1}{2}(M^{2}+\omega^{2})\,,
\\ \notag
\mathcal{L}_{0} & = & \frac{1}{2}\left[\partial_{\mu}y\partial_{\mu}y+\partial_{\mu}
\bit{z}\partial_{\mu}\bit{z}+\omega^{2}y^{2}+M^{2}\bit{z}^{2}\right]\,,
\\[2mm] \notag
\mathcal{L}_{1} & = & -ih(y^{2}+\bit{z}^{2})\partial_{0}y\,,
\\ \notag
\mathcal{L}_{2} & = & \frac{1}{2}(y\partial_{\mu}y+\bit{z}\partial_{\mu}\bit{z})
(y\partial_{\mu}y+\bar{z}\partial_{\mu}\bit{z})+\frac{\omega^{2}}{8}(y^{2}+\bit{z}^{2})^{2}\,.
\end{eqnarray}
Here $\mathcal{L}_{-2}$ is just a constant, $\mathcal{L}_{0}$ defines the kinetic term for the
two-dimensional fields $y(x)$ and $\bit{z}(x)$ while $\mathcal{L}_{1}$ and $\mathcal{L}_{2}$ define,
correspondingly, cubic and quartic interaction vertices. Notice that the Lorentz covariance of
$\mathcal{L}_{1}$ is broken by the external field.

Our goal is to calculate the change in the free energy density $\mathcal{F}(h)-\mathcal{F}(0)$
relative to its value for $h=0$. The free energy density is defined as
\be\label{F-def}
\e^{-V\mathcal{F}(h)}=\int\mathcal{D}y\mathcal{D}\bit{z}\,\e^{-\int
d^{D}x\,\mathcal{L}(x)}\,,\quad\qquad D=2-\epsilon\,.
\ee
It is ultraviolet divergent and we used dimensional regularization to define it properly.
Substitution of \re{L-dec} into \re{F-def} yields the perturbative expansion
\begin{align}\label{path}
\e^{-V\mathcal{F}(h)} &= \e^{\frac{M^{2}+m^{2}}{2\lambda^{2}}V}\int\mathcal{D}y\mathcal{D}\bit{z}
\e^{-\int d^{D}x\,\mathcal{L}_{0}(x)}
\\ \notag
& \times \left[1-\lambda^{2}\int d^{D}x\,\mathcal{L}_{2}(x) +\frac{\lambda^{2}}{2}\int
d^{D}x\,\mathcal{L}_{1}(x)\int d^{D}x'\mathcal{L}_{1}(x')+O(\lambda^{4})\right]\,.
\end{align}
Here the term linear in $\mathcal{L}_{1}$ is absent since it only involves odd powers of $y-$field
and, therefore, vanishes upon integration. Taking the logarithm of both sides of \re{path},
dividing by volume $V=\int d^D x$ and subtracting $\mathcal{F}(0)$ we get
\be\label{F-dec}
\mathcal{F}(h)-\mathcal{F}(0)=\frac{1}{\lambda^{2}}\mathcal{F}^{(-1)}(h)+  \mathcal{F}^{(0)}(h)
+\lambda^{2}\mathcal{F}^{(1)}(h)+O(\lambda^{4})\,.
\ee
We calculate each term separately for a
finite $\omega$ and put $\omega\to 0$ at the end of the calculation. We relegate the details of the
calculations to Appendix B and present here only the results.

The first term in the right-hand side of \re{F-dec} describes the classical contribution to the free
energy
\be
\mathcal{F}^{(-1)}(h)=-\frac{h^{2}}{2}\,.
\ee
The next order term, $\mathcal{F}^{(0)}(h)$, sums up the quadratic fluctuations of $y-$ and
$\bit{z}-$fields and it is related to the determinant of the kinetic operators in
$\mathcal{L}_0$. Using dimensional regularization it can be written as
\be\label{gamma}
\mathcal{F}^{(0)}(h) =\frac{n-2}{4\pi}h^{2-\epsilon} \left\{
\frac{1}{\epsilon}+\frac{\gamma}{2}+\frac{1}{2}\right\}\,,\quad \quad
\gamma=\Gamma^{\prime}(1)+\ln(4\pi)\,.
\ee
Finally, going through calculation of the the third term in the right-hand side of \re{F-dec} we
find
\be
\mathcal{F}^{(1)}(h) =\frac{n-2}{16\pi^{2}}h^{2-2\epsilon}\left\{
\frac{1}{\epsilon}+\gamma+\frac{1}{2}\right\}\,.
\ee
At this point, we combine together three terms and obtain the following expression for the free
energy density
\be\label{F-UV}
\mathcal{F}(h)-\mathcal{F}(0)=-\frac{h^{2}}{2\lambda^{2}}+\frac{n-2}{4\pi} h^{2-\epsilon}\left\{
\frac{1}{\epsilon}+\frac{\gamma}{2}+\frac{1}{2}\right\}
+\lambda^{2}\frac{n-2}{16\pi^{2}}h^{2-2\epsilon}\left\{
\frac{1}{\epsilon}+\gamma+\frac{1}{2}\right\} +O(\lambda^{4})\,.
\ee
The first two terms in this expansion were already calculated in \cite{HMN90}. The result for
the $O(\lambda^2)$ term is new and it is needed to make the comparison with the results of
Refs.~\cite{RT07} and \cite{G08}.

\subsection{Renormalization of the free energy density}

Ultraviolet divergences in the free energy \re{F-UV} can be removed in the standard way by
expressing the free energy in term of the renormalized coupling and using the renormalization group
to improve the result. In the $\overline{\rm MS}$ scheme, the relation between bare coupling
$\lambda$ and renormalized coupling $\tg(\mu)$ reads
\be\label{renorm}
\lambda^{2}\to\lr{\mu\e^{\gamma/2}}^{\epsilon}Z_{1}\tg^{2}\,,\quad\qquad
Z_{1}=1-\frac{2\beta_{0}\tg^{2}}{\epsilon}
-\frac{\beta_{1}\tg^{4}}{\epsilon}+\frac{4\beta_{0}^{2}\tg^{4}}{\epsilon^{2}}+\dots\,,
\ee
where $\gamma$ is defined in \re{gamma} and $\beta_{0,1,2}$ are the coefficients of the
$\beta-$function of the O($n$) model up to three loops
\be\label{beta1}
\mu\frac{d\tilde g}{d\mu}={\beta}(\tilde
g)=-\beta_{0}\tg^{3}-\beta_{1}\tg^{5}-\beta_{2}\tg^{7}+\dots
\ee
In the $\overline{\rm MS}$ scheme they have been determined in \cite{BH} to be
\begin{equation}\label{eq:Betafgv}
\beta_{0}=\frac{n-2}{4\pi}\,,\quad\qquad\beta_{1}=\frac{n-2}{8\pi^{2}}\,,\quad\qquad\beta_{2}^{(\overline{\rm
MS})} =\frac{(n+2)(n-2)}{64\pi^{3}}\,,
\end{equation}
where we introduced the superscript to indicate that $\beta_2$ is scheme-dependent.

Note that the coupling $\tg$  introduced here is analogous to the coupling $\bar g$ introduced in
Sect.~2.3. However the important difference between the two couplings is that they are defined in
two different schemes. Indeed, it is straightforward to verify that for $n=6$ the beta-functions
\re{beta} and \re{eq:Betafgv} coincide up to two loops, but they differ starting from three loops,
$\beta_{2}^{(\overline{\rm FRS})}\neq \beta_{2}^{(\overline{\rm MS})}$. Still,
the two couplings are related to each other through a finite renormalization.

With the relation \re{renorm} taken into account, the renormalized free energy density reads
\be\label{F-g}
\mathcal{F}(h)-\mathcal{F}(0)=-\frac{h^{2}}{2} \left\{ \frac{1}{\tg^{2}}
-2\beta_{0}\left(\ln\frac{\mu}{h}+\frac12\right)-2\beta_{1}\tg^{2}\left(\ln\frac{\mu}{h}
+\frac{1}{4}\right)+O(\tg^4)\right\}\,.
\ee
We verify with a help of \re{beta1} that the right-hand side of this relation does not depend on the
renormalization scale $\mu$. This suggests to express $\mathcal{F}(h)-\mathcal{F}(0)$ in a
renormalization group invariant way.

It is important to keep in mind that, contrary to the free energy density, the coupling $\tg(\mu)$
is not a physical quantity. We can explore this fact to define a new coupling constant to our best
convenience. The running of the coupling is determined by the Gell-Mann--Low equation (\ref{beta1})
and it depends on the scale $\Lambda_{\overline{\rm MS}}$ given by a general expression \re{Lam} in
the ${\overline{\rm MS}}$ scheme. Using this scale we define a new universal coupling $\alpha(h)$ as
\begin{equation}\label{aldef}
\frac{1}{\alpha}+\xi\ln\alpha=\ln\frac{h}{\Lambda_{\overline{\rm MS}}}\,,\qquad\quad
\xi=\frac{\beta_{1}}{2\beta_{0}^{2}}=\frac{1}{n-2}\,.
\end{equation}
Differentiating both sides of \re{aldef} with respect to $h$ we find the coupling $\alpha$ defines
the scheme in which beta-function is given to all loops by
\be
h \frac{d\alpha}{d h} =  \beta(\alpha) = -\frac{\alpha^2}{1-\xi\alpha}\,.
\ee
It has the advantage that any other coupling constant depends only polynomially on $\alpha$. In
particular, for $\mu=h$ we find from \re{beta1} and \re{aldef}
\be
\tg^2(h)  = \frac1{2\beta_0}\lr{\alpha +\frac{\xi}4 \alpha^3+O(\alpha^4)}\,.
\ee
Then, perturbative expansion of the free energy density \re{F-g} in $\alpha$ takes the form
\begin{equation} \label{F2}
\mathcal{F}(h)-\mathcal{F}(0)=-\beta_{0}h^{2}\left\{ \frac{1}{\alpha}-\frac{1}{2}
-\frac{\xi\alpha}{2}+O(\alpha ^2)\right\}\,.
\end{equation}
Here the first two terms reproduce the result of \cite{HMN90} and the third one is the sought next
order correction.

Let us now determine perturbative expansion of the energy density $\varepsilon(\rho)$. It can be
obtained from the free energy density \re{F2} through Legendre transformation
\be
\varepsilon(\rho)=\mathcal{F}(h)-\mathcal{F}(0)+\rho h\,,\qquad\qquad \rho=-\mathcal{F}^{'}(h)\,.
\ee
Using (\ref{aldef}) and (\ref{F2}) we find explicitly
\begin{align}\label{Leg}
\varepsilon &= \frac{\rho^{2}\alpha(h)}{4\beta_0}\left[
1+\frac{\alpha(h)}{2}+\frac{\xi\alpha^2(h)}{2}+O(\alpha^{2})\right],
\\ \notag
\rho &= 2\beta_{0}h\left[\frac{1}{\alpha(h)}+O(\alpha^{2})\right].
\end{align}
These relations define a parametric dependence of the energy density $\varepsilon$ on the particle
density $\rho$. To express $\varepsilon$ entirely as a function of $\rho$ we introduce yet another
coupling $\tilde{\alpha}(\rho)$ defined by
\be\label{a-tilde}
\frac{1}{\tilde{\alpha}}+(\xi-1)\ln\tilde{\alpha}=\ln\frac{\rho}{2\beta_{0} \Lambda_{\overline{MS}}}
\,. 
\ee
Replacing $\rho$ by its expression \re{Leg} and making use of \re{aldef}, we establish the relation
between the two couplings
\be
\frac{1}{\tilde{\alpha}(\rho)}+(\xi-1)\ln\tilde{\alpha}(\rho)=\frac{1}
{\alpha(h)}+(\xi-1)\ln\alpha(h) + O(\alpha^3)\,,
\ee
leading to $\alpha(h)= \tilde \alpha(\rho) + O(\tilde \alpha^4)$.

As a result, we obtain from \re{Leg} the energy density in the O($n$) model as
\begin{equation} \label{epsrho}
\varepsilon(\rho)= {\rho^{2}} \pi \xi \left\{  \tilde{\alpha}+\frac{\tilde{\alpha}^2}{2}
+\frac{\xi\tilde{\alpha}^{3}}{2}+O(\tilde{\alpha}^{4})\right\}\,.
\end{equation}
According to \re{a-tilde}, the coupling $\tilde\alpha$ depends on the scale $\Lambda_{\overline{\rm
MS}}$. To make a comparison with the string theory calculation we need its expression in terms of
the physical mass $m$. To this end, we take into account the known relation between the two
scales~\cite{HMN90}
\be
\zeta=\ln\frac{m}{\Lambda_{\widebar{\rm MS}}}=(3\ln 2-1)\xi-\ln\Gamma(1+\xi)\,,
\ee
and express the coupling \re{a-tilde} in terms of ${\rho}/{m}$ as
\begin{equation}\label{ujalfa}
\frac{1}{\tilde{\alpha}}+(\xi-1)\ln\tilde{\alpha} 
=\ln\frac{2\rho}{m}+A,\qquad A=\zeta+\ln\lr{\pi\xi}
\end{equation}
with $\xi=1/(n-2)$. 
Being combined together, the relations \re{epsrho} and \re{ujalfa} define the first few terms of the
perturbative expansion of the energy density in the O($n$) model in the perturbative regime $\rho\gg
m$.

\subsection{Numerical analysis of the TBA equation}

As shown in the second paper in \cite{HMN90}, the same function $\varepsilon(\rho)$ can be obtained
from the solution of the integral TBA equation (\ref{tba1}) in a parametric form given in
(\ref{tba2}). Using the generalized Wiener-Hopf technique one can transform equation (\ref{tba1}) to
a form more suitable both for finding the large $\rho$ asymptotic expansion of the energy density
and for numerical calculation. Employing this technique, Hasenfratz, Maggiore and Niedermayer
\cite{HMN90} calculated the first two terms in the large $\rho$ expansion of $\varepsilon(\rho)$ and
computed the ratio $m/\Lambda_{\rm \widebar{MS}}$.

Since for the calculation of the next order term we have to leave the beaten paths we decided to
check the result of the perturbation theory numerically for the O(6) model. The precision
measurement would require to perform several integral transformation, but a rough estimate can be
obtained directly form the original TBA equations (\ref{tba2}) and (\ref{tba1}) as follows: First
the parameter $B$ is fixed as an integer in the range $B=5,\dots,14$. On the interval $[-B,B]$ the
pseudo-energy $\chi(\theta)$ is discretized on $2^{18}\div 2^{24}$ points. Then the integral
equation (\ref{tba1}) is solved by iteration. (Technically, we used the fast Fourier transformation
routine of the numerical programming language octave to perform the convolution). From the solution,
$\chi(\theta,B)$, we calculated the ratios $\rho(B)/m$ and $\varepsilon(B)/m^2$ with a help of
(\ref{tba2}). Our relative precision for these quantities was as good as $10^{-5}$. Then we
calculated the coupling $\tilde{\alpha}(\rho)$ by numerically solving the equation (\ref{ujalfa}).
The results are displayed in Table \ref{table}.

\renewcommand{\baselinestretch}{1.5}
\begin{table}[!h]
\begin{tabular}{|c|c|c|c|c|c|}
\hline $B$ & 5 & 6 & 7 & 8 & 9\tabularnewline \hline $\tilde{\alpha}$ & 0.23355(3) & 0.19097(5) &
0.16135(3) & 0.13958(8) & 0.12293(8)\tabularnewline \hline $\rho/m$ & 94.8911(7) & 286.651(8) &
850.661(7) & 2492.50(2) & 7233.93(2)\tabularnewline \hline $\varepsilon/m^2$ & 1856.75(4) &
13560.90(5) & 99410.4(0) & 730337.7(8) & 5373044.(2)\tabularnewline  \hline \hline $B$ & 10 & 11 &
12 & 13 & 14\tabularnewline \hline $\tilde{\alpha}$ & 0.10979(9) & 0.099173(3) & 0.090405(5) &
0.083049(9) & 0.076792(3)\tabularnewline \hline $\rho/m$ & $2.08399(8)\cdot 10^4$ & $5.96823(1)
\cdot 10^4$ & $1.70101(3) \cdot 10^5$ & $4.82856(7)\cdot 10^5$ & $1.36601(2) \cdot 10^6$
\tabularnewline \hline $\varepsilon/m^2$ & $3.95663(6) \cdot 10^7$ & $2.91555(6) \cdot 10^8$ &
$2.14945(9) \cdot 10^9$ & $1.58524(9) \cdot 10^{10}$ & $1.16947(3) \cdot 10^{11}$\tabularnewline
\hline
\end{tabular}
\renewcommand{\baselinestretch}{1}%

\caption{Numerical result for the energy and particle density in the O(6) sigma model as function of
parameter $B$. Unreliable digits are displayed in parenthesis. } \label{table}
\end{table}
\renewcommand{\baselinestretch}{1}%

Finally we fitted the ratio ${\varepsilon}/\lr{\rho^2 \tilde{\alpha}}$ to the expression
\be\label{new}
\frac{\varepsilon}{\rho^2 \tilde{\alpha}} = \frac{\pi}4 \left( 1+\frac{\tilde{\alpha}}{2}
+\frac{\tilde{\alpha}^{2}}{8}+O(\tilde{\alpha}^{3})\right)\,,
\ee
which follows from (\ref{epsrho}) in the case of the O(6) model. Feeding in the known first two
coefficients we obtained $0.12(4)$ for the coefficient of the interesting $O(\tilde{\alpha}^{2})$
term in the right-hand side of \re{new}. This is in good agreement with the predicted value
$\frac{1}{8}$, recalling that here we are dealing with the coefficient of a second subleading
correction.

\begin{figure}[h!]
\psfrag{alpha}{$ \tilde \alpha$ } \psfrag{eps}[0][0][1][270]{$\displaystyle\frac{\varepsilon}{\rho
^2 \tilde \alpha }$}
\begin{centering}
\includegraphics[height=10cm]{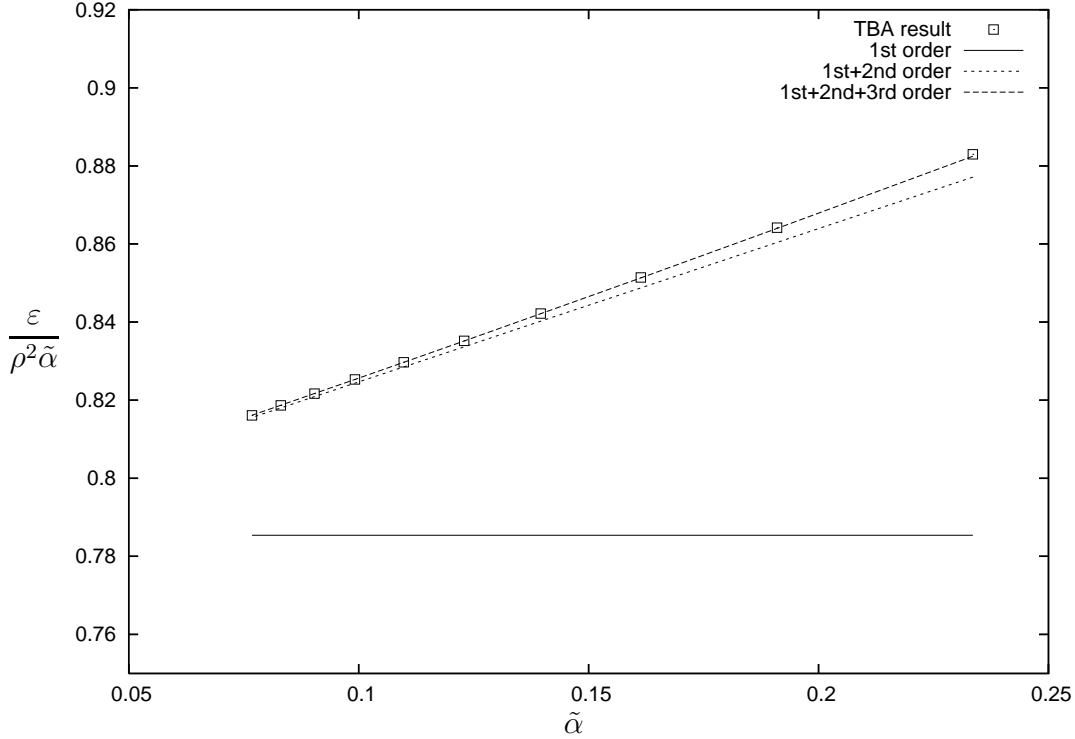}
\par\end{centering}

\caption{${\varepsilon}/\lr{\rho^{2}\tilde{\alpha}}$ is plotted against $\tilde{\alpha}.$ Boxes
represent the numerical TBA data  with invisible numerical errors. The lines correspond to
perturbative corrections as marked on the figure. } \label{tbafig}
\end{figure}

To visualize the result we collect the data on Figure~1, where we plot ${\varepsilon}/\lr{\rho^2
\tilde{\alpha}}$ against the coupling $\tilde{\alpha}$. Clearly the sum of all the three calculated
perturbative contributions approaches nicely the numerical solution giving a convincing confirmation
of both. We obtained similar result for the O$(3)$ model as well, thus confirming our perturbative
calculations numerically.

\subsection{Large $j$ expansion of the energy density}

To make a decisive comparison with the results of Refs.~\cite{RT07} and \cite{G08}, we have to
examine the behavior of the energy density $\varepsilon(\rho)$ for large particle density $\rho \gg
m$, or equivalently $j\gg m$ with the scaling variable $j=2\rho$ defined in \re{O6-epsilon}.

In this region, the coupling constant \re{ujalfa} is small and it has the form
\be\label{a-h}
\frac{1}{\tilde{\alpha}}=\ln\frac{j}{m}+a_1\ln\ln\frac{j}{m}+a_2+
a_3\frac{\ln\ln\frac{j}{m}}{\ln\frac{j}{m}}+\frac{a_4}{\ln\frac{j}{m}}+\dots\,.
\ee
The coefficients $a_i$ can be determined recursively from (\ref{ujalfa}) leading to
\be
a_1=\xi-1,\qquad a_2=A,\qquad a_3=(\xi -1)^2,\qquad a_4=(\xi -1)A\,,
\ee
with the constants $\xi$ and $A$ defined in \re{aldef} and \re{ujalfa}, respectively. We substitute
the relation \re{a-h} into (\ref{epsrho}) and obtain, after some algebra, the following expression for
the energy density of the O($n$) model
\begin{align}\label{e-general}
\varepsilon_{{\rm O}(n)} =\frac{j^2\pi \xi}{4\ln\frac{j}{m}}\Bigg\{ 1 & +
\frac{(1-\xi)}{\ln\frac{j}{m}}\left[\ln\left( {\kappa}\ln\frac{j}{m}\right)+\frac{1}{2}\right]
\\ \notag
 & +\frac{(1-\xi)^2}{\ln^2\frac{j}{m}}\left[\ln^2\left( {\kappa}\ln\frac{j}{m}\right)+
\frac{\frac{\xi}{2}\left(1-\frac{\xi}{2}\right)}{(1-\xi)^2}\right]+\dots \Bigg\}\,,
\end{align}
where the notation was introduced for
\be\label{const}
\ln {\kappa}=\frac{\frac{\xi}{2}-A}{1-\xi}=\frac{\lr{\ft32-3\ln
2}\xi+\ln\lr{\Gamma(\xi)/\pi}}{1-\xi} \,.
\ee
We recall that the parameter $j=2\rho$ defines the density of particles with mass $m$ and the
relation \re{e-general} only holds for $j\gg m$.

We are now ready to perform a comparison of the energy density of the O(6) model and the scaling
function \re{e-pt} in the AdS/CFT. For $n=6$ the constants \re{const} take the following values
\be
 \xi =\frac{1}{4}\,,\qqqquad
\ln {\kappa}=\frac{1}{2}-\frac{1}{3}\ln2-\frac{4}{3}\ln\Gamma\bigl(\ft{3}{4}\bigr) \,.
\ee
Taken into account these relations we find that, in agreement with \re{O6-epsilon}, the expression
for $2\varepsilon_{\rm O(6)}$ indeed coincides with the two-loop result for the scaling function
(\ref{e-pt}) with
\be\label{e1}
\epsilon_1=\frac{\frac{\xi}{2}(1-\frac{\xi}{2})}{(1-\xi)^2}\bigg|_{n=6}=\frac{7}{36}\,.
\ee
Finally, we substitute the relations (\ref{megy}) and \re{e1} into \re{q02} and compute the constant
$q_{02}$ as
\be
q_{02}=\frac{9}{8}+\frac{3}{4}-\frac{3}{2}\ln2+\frac{9}{2}\cdot\frac{7}{36}=
\frac{11}{4}-\frac{3}{2}\ln2\,.
\ee
Comparing this relation with \re{discrepancy} we conclude that our result for $q_{02}$ is in
agreement with the result obtained from the quantum string Bethe ansatz~\cite{G08}.

\section{Conclusions}

In this paper, we applied methods of integrable models previously developed for the
two-dimen\-sional O($n$) sigma-model to study the scaling function in the AdS/CFT. The starting point
of our analysis were the relations \re{O6-epsilon} and \re{m_AM} which relate the energy
density of the O(6) model to the scaling function and identify the mass gap of the O(6)
model $m$ with a new dynamical scale in the AdS/CFT.

These relations are extremely nontrivial given the fact that the scale $m$ has a different origin in
the models under consideration. The O(6) model is asymptotically free at short distances and the
mass scale arises due to a nontrivial dynamics in the infrared. At the same time, $\mathcal{N}=4$
SYM and string sigma-model on $\rm AdS_5\times S^5$ are conformal invariant theories and, therefore,
they do not generate any scale. In the string theory, the scaling function describes quantum
corrections to the folded string spinning on $\rm AdS_5\times S^5$ and it is the underlying
classical configuration that introduces the mass scales. In a similar manner, in $\mathcal{N}=4$ SYM
theory the scaling function follows from the analysis of the Bethe ansatz equations in the limit
\re{j}. In this limit, the Bethe roots condense on the real axis and their distribution depends on
the parameters \re{j} fixed by the quantum numbers of Wilson operators. It is therefore remarkable
that our calculation of the mass scale in $\mathcal{N}=4$ SYM theory reproduces the known result for
the mass gap in the O(6) model in the special renormalization scheme dictated by the FRS equation.

For $g\to\infty$ and $j/m=\rm fixed$, the scaling function can be found exactly by solving the TBA
equations for the energy density of the O(6) model. Finding solution to these equations in the
perturbative regime $j\gg m$ beyond the leading order proves to be a difficult task. We demonstrated
that the problem can be circumvented by employing methods of standard perturbation theory to
calculating the free energy of the O($n$) model in the presence of an external field. In this way,
we computed the two-loop correction to the energy density of the O($n$) model and verified that it
agrees with the numerical solution to the TBA equation. Then, we combined this correction (for
$n=6$) with the two-loop correction to the mass scale $m$ and calculated the scaling function.

Comparing the obtained expression with two different predictions for the scaling function coming
from two-loop stringy corrections to the folded spinning string solution~\cite{RT07} and from
thermodynamical limit of quantum string Bethe ansatz equations~\cite{CK07,B08,G08}, we found an
agreement with the latter one. The agreement should not be surprising since the FRS equation was
originally obtained from the all-loop Bethe ansatz for the dilatation operator in ${\cal N}=4$ SYM
theory in an appropriate scaling limit.

The question remains however what is the reason for a disagreement between our result for the
scaling function and the explicit two-loop stringy calculation. Notice that the difference only
amounts to a constant term (see Eqs.~\re{e-2-loop} and \re{discrepancy}) while logarithmically
enhanced terms coincide. This implies that the two-loop stringy result is consistent with the
relation \re{O6-epsilon} between the scaling function and the energy density of the O(6) model but
it leads to the expression for the two-loop correction to the mass scale \re{m1} which is different
from \re{megy},
 \be
m_1^{\rm str} = -\frac1{32}-\frac3{16} \ln 2 -\frac14\,{\rm K}\,.
\ee
This suggests that either the all-loop Bethe ansatz does not predict correctly the mass scale, or
the two-loop stringy calculation needs to be revisited.
 At present stage we can not discriminate
between these two scenarios and the question requires further investigation.~\footnote{A potential  difficulty in comparing the two predictions is that the string computation \cite{RT07}  was performed in a scheme in which two-loop beta-function coefficient is zero. This scheme
is related to the $\widebar{\rm MS}$ scheme by a singular coupling redefinition and it was previously  used \cite{scheme} in the studies of two-dimensional bosonic O($n$) model. We are grateful to A.~Tseytlin for drawing our attention to this fact.}

\section*{Acknowledgments}

We would like to thank Changrim Ahn, Alexander Gorsky, Juan Maldacena, Radu Roiban, Mikhail Shifman and Arkady
Tseytlin for interesting discussions. ZB, JB and LP thank the APCTP, Focus Program ``Finite-size
technology in low dimensional quantum field theory", for the hospitality in Pohang where part of the
work was performed. ZB an LP thank for a partial financial support of OTKA K60040.  JB was partially
supported by the OTKA grant T-049495. ZB was supported by a Bolyai Scolarship and the EC network
``Superstring''. BB and GK were supported in part by the French Agence Nationale de la Recherche
under grant ANR-06-BLAN-0142.

\appendix

\setcounter{section}{0} \setcounter{equation}{0}
\renewcommand{\theequation}{A.\arabic{equation}}

\section*{Appendix A:\ \ Calculation of the mass scale}

Let us now compute the mass gap \re{m=int}. At large $g$ the integral in \re{m=int} receives a
dominant contribution from $t\sim g$. In order to evaluate \re{m=int} it is convenient to change the
integration variable as $t\to 4\pi g i t$. Then, we get from \re{m=int}
\be\label{m=int1}
m =  \frac{8\sqrt{2}}{\pi^2}\e^{-\pi g}+\Delta m\,,
\ee
where $\Delta m$ is given by integral of $\Gamma(4\pi g it)$ along the imaginary axis
\begin{align}\label{i1}
\Delta m = -\frac{4g}{\pi}\e^{-\pi g}  \int_0^{-i\infty}  \frac{dt\, \e^{-4\pi g t-i\pi/4}}
{t+\ft14}\lr{\Gamma^{\BES}_{+}(4\pi g it)+ i \Gamma^{\BES}_{-}(4\pi g it) } +\text{c.c.}
\end{align}
According to \re{G-final} the functions $\Gamma_\pm(4\pi g it)$ have the following form
\be\label{i2}
\Gamma^{\BES}_{+}(4\pi g it)+ i \Gamma^{\BES}_{-}(4\pi g it)  =  V_0(4\pi g t) f_0(t) +  V_1(4\pi g
t) f_1(t)\,,
\ee
where $f_0$ and $f_1$ are defined as coefficients in front of $V_0$ and $V_1$, respectively, in the
right-hand side of \re{G-final}. They are given by a linear combination of the ratio of Euler
gamma-functions
\begin{align} \notag
f_0(t) &= - \frac{{\Gamma}(\ft34)\Gamma(1-t)}{\Gamma(\ft34-t)}+ \frac{1}{4\pi g}\left[\lr{\frac{3\ln
2}{4}+\frac1{8t}} \frac{\Gamma(\ft34)\Gamma(1-t)}{\Gamma(\ft34-t)}-
\frac{\Gamma(\ft14)\Gamma(1+t)}{8t\, \Gamma(\ft14+t)} \right]+O(g^{-2})\,,
\\ \label{ff}
f_1(t) &= \frac{1}{4\pi
g}\left[\frac{\Gamma(\ft14)\Gamma(1+t)}{4t\,\Gamma(\ft14+t)}-\frac{\Gamma(\ft34)\Gamma(1-t)}{4t\,\Gamma(\ft34-t)}
\right]+O(g^{-2})\,.
\end{align}
Notice that $f_1(t)$ is suppressed by factor $1/(4\pi g)$ compared to $f_0(t)$.

To work out the large $g$ expansion of the integral \re{i1} it is convenient to use the
Mellin-Barnes representation for the functions $V_0(4\pi gt)$ and $V_1(4\pi gt)$. Using the integral
representation \re{II} we find
\begin{align}
V_0(4\pi gt)\e^{-4\pi g t} =
\frac{\sqrt{2}}{\pi}\int_{-\delta-i\infty}^{-\delta+i\infty}\frac{dj}{2\pi i}\Gamma(-j) (4\pi g t)^j
\int_{-1}^1 du \, (1-u)^{j-1/4} (1+u)^{j+1/4}
\end{align}
and similar representation also exists for $V_1(4\pi gt)\e^{-4\pi g t}$. Their substitution into
\re{i2} and \re{i1} yields after $u-$integration
\begin{align}\label{r1}
\Delta m =  -\frac{4\sqrt{2}g}{\pi^2}\e^{-\pi g} \Gamma(\ft14)  &
\int_{-\delta-i\infty}^{-\delta+i\infty}\frac{dj}{2\pi
i}\frac{\Gamma(-j)\Gamma(j+\ft34)}{\Gamma(j+2)} (2\pi g)^j  \e^{-i\pi/4}
\\ \notag  \times &
\int_0^{-i\infty}  \frac{dt\,(4t)^j}{t+\ft14}\left[ \frac12 {f_0(t)} + {f_1(t)} (j+1) \right]
  +\text{c.c.}
\end{align}
To find the asymptotic expansion at large $g$ we deform the integration contour in the complex $j$
plane to the left and pick up the contribution of poles at negative $j$. These poles come  from
$\Gamma(j+\ft34)$ and from $t-$integral. Let us start with the latter ones.

By definition,  $f_0(t)$ and $f_1(t)$  are real meromorphic functions of $t$ with simple poles
located at $t=\pm 1, \pm 2, \ldots$. Then,  integration over small $t$ produces simple poles located
at negative integer $j$. However, due to the presence of $1/\Gamma(j+2)$, all these poles except
$j=-1$ produce vanishing contribution to the $j-$integral. Calculating the residue at $j=-1$ we find
\begin{align}\label{d1}
\Delta m =  -\frac{4\sqrt{2}g}{\pi^2}\e^{-\pi g} \Gamma(\ft14)  &  { \Gamma(-\ft14)} (2\pi g)^{-1}
\e^{-i\pi/4}\frac1{2} {f_0(0)}    +\text{c.c.}  + \ldots = -\frac{8\sqrt{2}}{\pi^2}\e^{-\pi g} +
\ldots\,,
\end{align}
where ellipses denote the contribution of poles produced by $\Gamma(j+\ft34)$ in \re{r1}. We notice
that \re{d1} cancels against similar term in the right-hand side of \re{m=int1} and, therefore, $m$
is determined by the contribution of poles at $j=-\ft34,-\ft74,\ldots$. Going through calculation of
residues we obtain
\begin{align}
m = \frac{4ig}{\pi^2} \Gamma(\ft34)(2\pi g)^{-3/4}\e^{-\pi g}\int_{-\delta-i\infty}^{-\delta+i\infty}
\frac{dt\,(- t)^{-3/4} } {t+\ft14} \left[ f_0(t) + \frac12 f_1(t) - (4\pi g t)^{-1}\frac3{32}
f_0(t) +O(g^{-2})\right],
\end{align}
where the integration contour goes to the left from the branch cut that starts at $t=0$.
Calculating this integral and taking into account analytical properties of the functions \re{ff} we
find
\begin{align}
m & = - \frac{4g}{\pi}(2\pi g)^{-3/4}\e^{-\pi g}\Gamma(\ft34)2^{5/2} \left[f_0(-\ft14) + \frac12 f_1(-\ft14)
+\frac3{32 \pi g} f_0(-\ft14) + O(g^{-2})\right]\,,
\end{align}
leading to
\begin{align}\label{m-f}
m & =  (2\pi g)^{1/4}\e^{-\pi g} \frac{ 2^{1/2}}{\Gamma(\ft54)}\left[ 1 - \frac{6\ln 2 -3}{32\pi
g}+O(g^{-2})\right],
\end{align}
in an agreement with \re{m}.

The strong coupling expansion of the mass scale \re{m-f} can be systematically
improved by
taking into account subleading $1/g$ corrections to the functions $f_0(t)$ and $f_1(t)$ in \re{ff}. Assuming that
higher order corrections do not modify analytical properties of these functions and taking into account
a contribution to \re{r1} from an infinite sequence of poles at $j=-3/4-n$ (with $n$ nonnegative integer)
we find after some algebra
\begin{align}
m & = - \frac{4g}{\pi}\e^{-\pi g}2^{5/2} \left[f_0(-\ft14) U_0^-(\pi g)+   f_1(-\ft14)U_1^-(\pi g)
 \right]\,.
\end{align}
Here the notation was introduced for the functions
\begin{align}\notag
U_0^-(y) &= \int_0^\infty dt\, \e^{-2y t} t^{-1/4} (1+t)^{1/4}
 =
(2y)^{-3/4} \Gamma(\ft34) \left[ 1 +\frac3{32 y}+\ldots\right],
\\
U_1^-(y) &= \frac12 \int_0^\infty dt\, \e^{-2y t} t^{-1/4} (1+t)^{-3/4}
= (2y)^{-3/4} \ft12\Gamma(\ft34)\left[ 1
-\frac{9}{32 y} +\ldots \right],
\end{align}
which can be expressed  in terms of Whittaker functions of second kind
\begin{align} \notag
U_0^-(y) &= \ft12 \Gamma(\ft34) y^{-1} \e^y W_{1/4,1/2}(2y)\,,
\\[2mm]
U_1^-(y) &= \ft12 \Gamma(\ft34) (2y)^{-1/2} \e^y W_{-1/4,0}(2y)\,.
\end{align}

\setcounter{section}{0} \setcounter{equation}{0}
\renewcommand{\theequation}{B.\arabic{equation}}

\section*{Appendix B:\ \ Perturbative calculation of the free energy density}

In this appendix we calculate the constant (${\cal O}(\lambda^0)$) and quadratic (${\cal
O}(\lambda^2)$) terms in the perturbative expansion of the free energy \re{F-dec}.

The constant term sums up the quadratic fluctuations of fields $y(x)$ and $\bit{z}$ and it is given
by a logarithm of the ratio of determinants of the kinetic operators:
\be
\mathcal{F}^{(0)}(h)=\frac{n-2}{2V}\left[\mathrm{Tr}
\log(-\partial^{2}+M^{2})-\mathrm{Tr}\log(-\partial^{2}+\omega^{2})\right]\,,
\ee
with $M^2=h^2+\omega^2$ and $V$ being the volume factor. Fortunately computing the difference also
provides a regularization. By differentiating and integrating with respect to $M$ and $\omega$ we
can cast the result into the form
\be\label{F0}
\mathcal{F}^{(0)}(h)=\frac{n-2}{2}\lim_{\omega^{2}\to0}
\int_{\omega^{2}}^{M^{2}}dm'^{2}\int\frac{d^{D}p}{(2\pi)^{D}}\frac{1}{p^{2}+m'^{2}}\,.
\ee
The momentum integration can be performed in the dimensional regularization with $D=2-\epsilon$ as
\be\label{I}
I(m)=\int\frac{d^{D}p}{(2\pi)^{D}}\frac{1}{p^{2}+m^{2}}=
m^{D-2}\frac{\Gamma(1-\frac{D}{2})}{(4\pi)^{\frac{D}{2}}}=
\frac{m^{-\epsilon}}{4\pi}\left[\frac{2}{\epsilon}+\gamma+O(\epsilon)\right]\,,
\ee
with $\gamma=\Gamma^{\prime}(1)+\ln(4\pi)$. As we will see in a moment, the same integral appears in
higher order calculations. Using \re{I}, we find for the constant term \re{F0} (up to corrections
vanishing as $\epsilon\to 0$)
\be
\mathcal{F}^{(0)}(h)=\frac{n-2}{4\pi}h^{2-\epsilon} \left\{
\frac{1}{\epsilon}+\frac{\gamma}{2}+\frac{1}{2}\right\}\,.
\ee
The ${\cal O}(\lambda^2)$ term in the expansion \re{F-dec} describes the two-loop correction to the
free energy. It receives contribution from the last two terms inside the square brackets in
\re{path}
\be\label{F1/V}
\mathcal{F}^{(1)}(h)= \frac1{V}\int d^{D}x\,\vev{\mathcal{L}_{2}(x)}_0 - \frac{1}{2V}\int d^{D}x\int
d^{D}x'\,\vev{\mathcal{L}_{1}(x)\mathcal{L}_{1}(x')}_0\,,
\ee
with $\mathcal{L}_{1}(x)$ and $\mathcal{L}_{2}(x)$ defined in \re{L's}. Here the subscript `(0)'
indicates that the expectation values are evaluated with the measure
$\int\mathcal{D}y\mathcal{D}\bit{z} \exp\lr{-\int d^{D}x\,\mathcal{L}_{0}(x)}$ (see Eqs.~\re{path}
and \re{L's}).

The expression for ${V}^{-1}\int d^{D}x\,\vev{\mathcal{L}_{2}(x)}_0=\vev{\mathcal{L}_{2}}_0$ can
be obtained in terms of the two VEV's
\be \langle
y^{2}\rangle_{0}=I(\omega)\,,\quad\qquad\langle\bit{z}^{2}\rangle_{0}=(n-2)I(M)\,.
\ee
Using translational invariance $\langle\partial^{2}(y^{4})\rangle_{0}=0$ and the equation of motion
$\partial^{2}y=\omega^{2}y$, we get $\langle(y\partial_\mu
y)^{2}\rangle_{0}=-\frac{\omega^{2}}{3}\langle y^{4}\rangle_{0}=-\omega^{2}\langle
y^{2}\rangle_{0}^{2}$. Making use of similar relations for $\bit{z}-$field together with the
factorization property $\langle  y\partial_\mu y \bit{z}\partial_\mu\bit{z}\rangle_{0} =\langle
y\partial_\mu y\rangle_{0} \langle\bit{z}\partial_\mu\bit{z}\rangle_{0}=0$, the contribution of
$\mathcal{L}_{2}$ to \re{F1/V} can be written in the following way:
\begin{eqnarray}
\langle\mathcal{L}_{2}\rangle_{0} & = &
-\frac{\omega^{2}}{2}I^{2}(\omega)-\frac{(n-2)M^{2}}{2}I^{2}(M)
\\ \notag
 &  &
 +\frac{\omega^{2}}{8}\left[3I^{2}(\omega)+2(n-2)I(\omega)I(M)+2(n-2)I^{2}(M)+(n-2)^{2}I^{2}(M)\right]\,.
 \end{eqnarray}
We remove the infrared cut-off by taking the $\omega\rightarrow0$ limit and obtain the relevant
contribution from $\mathcal{L}_{2}$ as
\be\label{L2/V}
\frac1{V}\int
d^{D}x\,\vev{\mathcal{L}_{2}(x)}_0=\langle\mathcal{L}_{2}\rangle_{0}=-\frac{n-2}{2}h^{2}I^{2}(h)\,.
\ee
The contribution to the free energy \re{path} quadratic in $\mathcal{L}_{1}(x)$ has the form
\be\label{L1L1}
\frac{1}{2V}\int d^{D}x \int d^{D}x'\langle\mathcal{L}_{1}(x)
\mathcal{L}_{1}(x')\rangle_{0}=\frac{1}{2}\int d^{D}x\,
\langle\mathcal{L}_{1}(x)\mathcal{L}_{1}(0)\rangle_{0}\,.
\ee
By writing
\be
\mathcal{L}_{1}=-ih(y^{2}+\bit{z}^{2})\partial_{0}y=-ih\left(\ft13\partial_{0}{y^{3}}+\bit{z}^{2}
\partial_{0}y\right)\,,
\ee
we can see that the first term, being a total derivative, can be dropped. The remaining term gives
rise to the diagrams shown on Fig. \ref{diagfig}.
%
\newcommand{\inclfig}[2]{\mbox{\epsfysize=#1cm \epsfbox{#2.ps}}}%
\newcommand{\insertfig}[2]{\mbox{\epsfysize=#1cm \epsfbox{#2.eps}}}%
\begin{figure}[h]%
\begin{center}%
\mbox{\begin{picture}(0,60)(150,0)%
\psfrag{x}[rc][cc]{$x$} \psfrag{q}[cc][cc]{$q$} \psfrag{p}[cc][cc]{$p$}%
\psfrag{r}[cc][cc]{$r$}\psfrag{0}[cc][cc]{$0$} \put(0,0){\insertfig{1.5}{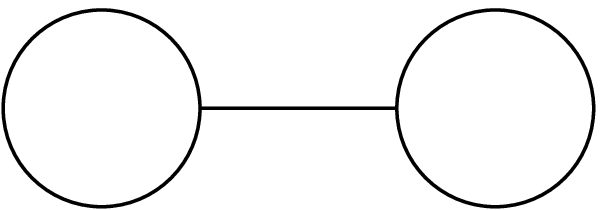}} \put(46,10){$x$}%
\put(74,9){$0$} \put(200,0){\insertfig{1.95}{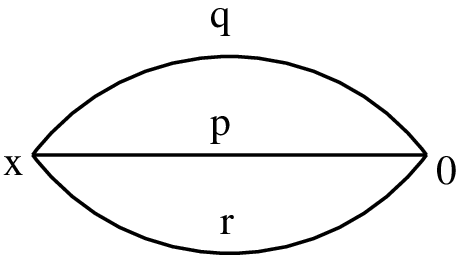}}%
\end{picture}}%
\end{center}%
\caption{\small Two-loop diagrams contributing to \re{L1L1}. Solid lines denote two-dimensional scalar propagators.}%
\label{diagfig}
\end{figure}%
%
The contribution of the first diagram on Fig. \ref{diagfig} involves the factor $\sim\int d^D
x\,\langle\partial_{0}y(x)\partial_{0}y(0)\rangle_0$ with the integrand being a total derivative
again. Thus the right-hand side of \re{L1L1} only receives contribution from the second diagram on
Fig \ref{diagfig}. Taking into account all possible contractions one finds
\be\label{trick}
\frac12\int d^{D}x\langle\mathcal{L}_{1}(x)\mathcal{L}_{1}(0)\rangle_{0}= -h^{2} (n-2)\int
d^{D}x\int\frac{d^{D}q}{(2\pi)^{D}}\frac{\e^{iqx}}{q^{2}+
M^{2}}\int\frac{d^{D}p}{(2\pi)^{D}}\frac{p_{0}^{2}\e^{ipx}}{p^{2}+\omega^{2}}
\int\frac{d^{D}r}{(2\pi)^{D}}\frac{\e^{irx}}{r^{2}+M^{2}}\,,
\ee
where the factor $(n-2)$ counts the number of $\bit{z}-$fields circulating inside the loop. Doing
the $p-$integration we observe that despite the fact that the integrand is not Lorentz covariant,
the integral in the right-hand side of \re{trick} could only depend on $M^2$ and $\omega^2$ and,
therefore, it should be Lorentz invariant. This allows us to simplify the $p-$integral as:
\be
\int\frac{d^{D}p}{(2\pi)^{D}}\frac{p_{0}^{2}\e^{ipx}}{p^{2}+\omega^{2}} \quad \Longrightarrow\quad
\frac{1}{D} \int\frac{d^{D}p}{(2\pi)^{D}}\frac{p^{2}\e^{ipx}}{p^{2}+\omega^{2}}=
\frac{1}{D}\delta^{(D)}(x)+O(\omega\ln \omega)\,.
\ee
As a consequence, the $x-$integral in \re{trick} becomes trivial and the remaining momentum
integration gives $I^2(M)$. Then, taking the $\omega\to0$ limit, we find from \re{trick}
\be\label{LL1/V}
\frac{1}{2}\int
d^{D}x\,\langle\mathcal{L}_{1}(x)\mathcal{L}_{1}(0)\rangle_{0}=-\frac{h^{2}}{D}(n-2)I^{2}(h)\,.
\ee
Substituting the relations \re{L2/V} and \re{LL1/V} into \re{F1/V} we derive the sought second order
correction to the free energy density
\be
\mathcal{F}^{(1)}(h)=h^{2}(n-2)I^{2}(h)\left[\frac{1}{D}-\frac{1}{2}\right]
=\frac{n-2}{16\pi^{2}}h^{2-2\epsilon}\left\{ \frac{1}{\epsilon}+\gamma+\frac{1}{2}\right\}\,.
\ee

\end{document}